\documentclass{emulateapj}
\usepackage{apjfonts}

\def\mic{$\mu$m}

\newcommand{\mum}{\ifmmode{\rm \mu m}\else{$\mu$m}\fi}

\shorttitle{The IR spectrum of SMP~LMC~11}
\shortauthors{Bernard-Salas et al.}

\begin{document}

\title{The Spitzer-IRS spectrum of  SMP~LMC~11}

\author{J. Bernard-Salas\altaffilmark{1}, E. Peeters\altaffilmark{2},
  G.C. Sloan\altaffilmark{1}, J. Cami\altaffilmark{2}, S.
  Guiles\altaffilmark{1}, J.R. Houck\altaffilmark{1}}

\altaffiltext{1}{Center for Radiophysics and Space Research, Cornell
  University, 219 Space Sciences Building, Ithaca, NY 14853-6801, USA.
  jbs@isc.astro.cornell.edu, sloan@isc.astro.cornell.edu,
  jrh13@cornell.edu}
\altaffiltext{2}{SETI Institute, 515 N. Whisman Drive, Mountain View,
  CA 94043. epeeters@mail.arc.nasa.gov, jcami@mail.arc.nasa.gov}

\begin{abstract}
  
  We present the first mid-infrared spectra of SMP~LMC\,11 in the Large
  Magellanic Cloud. While this object resembles a planetary nebula in
  the optical, its infrared properties are more similar to an object
  in transition from the asymptotic giant branch to the planetary
  nebula phase.  A warm dust continuum dominates the infrared
  spectrum. The peak emission corresponds to a mean dust temperature
  of 330~K. The spectrum shows overlapping molecular absorption bands
  from 12 to 17\,$\mu$m corresponding to acetylene and polyacetylenic
  chains and benzene.  This is the first detection of C$_4$H$_2$,
  C$_6$H$_2$, C$_6$H$_6$ and other molecules in an extragalactic
  object.  The infrared spectrum of SMP~LMC~11 is similar in many ways
  to that of the pre-planetary nebula AFGL~618. The IRS spectrum shows
  little evidence of nitrogen-based molecules which are commonly seen
  in Galactic AGB stars. Polycyclic aromatic hydrocarbons are also
  absent from the spectrum.  The detection of the \ion{[Ne}{2]}\,12.8
  $\mu$m line in the infrared and other forbidden emission lines in
  the optical indicates that an ionized region is present.

\end{abstract}


\section{Introduction}

The transition from the asymptotic giant branch (AGB) to the planetary
nebula (PN) phase remains an elusive stage in stellar evolution.  At
the end of its lifetime on the AGB, a star loses mass at high rates,
triggering the formation of molecules and dust in a thick
circumstellar envelope \citep[e.g.][]{Habing96,Blommaert:ISOlegacy}.
Radiation pressure on the dust and coupling between the dust and gas
drive the outflows, stripping the star and exposing its hot core.  As
the effective temperature of the exposed core rises, it will ionize
the nebular gas.

The detailed chemistry involved in this brief transition phase is
poorly understood.  In carbon-rich stars, carbon is primarily
incorporated first into carbon monoxide (CO) and then into acetylene
(C$_2$H$_2$).  Acetylene and its derivatives are the precursors from
which carbon-based compounds such as polycyclic aromatic hydrocarbons
(PAHs) and soot are formed \citep[][]{atb89,ff89}.  Here, we present
the infrared spectrum of SMP~LMC~11, an object in the Large Magellanic
Cloud (LMC) whose infrared spectral characteristics fall in the stage
where most of this complex chemistry takes place.

SMP~LMC~11 was part of the \citet{san} survey of PNe in the Magellanic
Clouds (from which it takes its designation, SMP~LMC~11).  It was also
included in the catalogue by \citet{lei} of accurate positions of
known PNe in the LMC.  The reported H$\beta$ flux and extinction vary
among papers in the literature. For instance, \cite{lei2}, \cite{woo}
and \cite{mea} report a log(H$\beta$)=$-$13.14 (with H$\beta$ in units
of erg cm$^2$ s$^{-1}$), whereas \citet{sha2} obtain
log(H$\beta$)=$-$13.94.  Similarly, differences in visual extinction
amount to a factor of more than 3.  The presence of another source
nearby in the IRAC images (see \S2) may explain these discrepancies.
\citet{mor} presented optical spectroscopy of 97 faint PNe including
SMP~LMC~11 and identified the forbidden emission lines of
\ion{[O}{3]}, \ion{[N}{2]} and \ion{[S}{2]}. Recently \citet{sha2}
observed it with the {\it Hubble Space Telescope} ({\it HST}) as part
of their sample of objects in the LMC and SMC. They measured the
strength of the \ion{[O}{1]} line, along with the previously
identified forbidden lines, and inferred a bipolar morphology with a
size of 0.76\arcsec$\times$0.55\arcsec.

In this letter we report the first mid-infrared spectrum of SMP
LMC\,11.  The spectrum reveals a wealth of molecular absorption bands
and a few atomic emission lines.  To our knowledge, the only other
object showing a similar infrared spectrum is the Galactic
pre-planetary nebula\footnote{We adopt the term ``pre-planetary
  nebula'' to avoid the confusing term ``protoplanetary nebula,''
  which could imply an object evolving into a PN or a nebula with
  embedded protoplanets.}  (PPN) AFGL~618.  \citet{buj} inferred that
AFGL~618 became a PPN only 200\,yr ago. \cite{chi98} identified simple
aliphatics (CH$_2$, CH$_3$) in its spectrum at 3.4 $\mu$m.
\cite{cer01a} studied the spectrum of AFGL~618 from the
Short-Wavelength Spectrometer \citep[SWS,][]{deg}  on the {\it
  Infrared Space Observatory} \citep[{\it ISO,}][]{kes}, and they
describe the system as having a thick molecular envelope surrounding a
B0 star and an ultra-compact H\,II region.  They suggest that UV
radiation from the central star and shocks in the high-velocity winds
have significantly modified the chemistry.  \cite{cer01a} detected
C$_4$H$_2$, C$_6$H$_2$ and of C$_6$H$_6$ in a circumstellar
environment for the first time.  In this letter we compare the
observed features in SMP~LMC~11 to those of AFGL~618.

The next section describes the observations and the data reduction
techniques.  \S 3 gives the analysis of the observed spectral features
and compares SMP~LMC~11 to AFGL~618.  \S 4 discusses the evolutionary
status of the object.

\section{Observations and data reduction}
\label{obs_s}

We observed SMP~LMC~11 with the Infrared Spectrograph
\citep[IRS,][]{hou} on board the {\it Spitzer Space Telescope}
\citep{wer} as part of the GTO program on 2005 June 6 (Program ID 103,
AORkey 4947712).  These observations consist of spectra from all four
IRS modules: Short-Low (SL), Long-Low (LL), Short-High (SH), and
Long-High (LH).  Table 1 gives the wavelength coverage, spectral
resolution and integration times for each module. We performed peak-up
on a nearby star to achieve accurate pointing (0.4$\arcsec$).

\begin{deluxetable}{c c c c c}
  \tablewidth{8cm} \tablecaption{Observation Log of
    SMP~LMC~11.\label{log_t}} \tablehead{\colhead{Module} & \colhead{Order}
    & \colhead{Wavelength (\mic)} & \colhead{Resolution} &
    \multicolumn{1}{c}{Obs. time\tablenotemark{a}}}
    \startdata
    SL &  1        &  7.5--14.5   &  60-120 &  14$\times$2   \\ 
    SL &  2        &  5.0--7.5    &  60-120 &  14$\times$2   \\ 
    LL &  1        &  20--40      &  60-120 &  30$\times$1   \\ 
    LL &  2        &  14.5--20    &  60-120 &  30$\times$1   \\ 
    SH &  11 to 20 &  10--20      &  600    &  60$\times$1   \\ 
    LH &  11 to 20 &  20--36      &  600    &  120$\times$1  \\ 
    \enddata
    \tablenotetext{a}{On-source observation time in seconds $\times$
      number of cycles.}
\end{deluxetable}

The data were processed through a copy of the {\it Spitzer} Science
Center's pipeline reduction software (version S13.2) maintained at
Cornell.  To avoid possible problems in the flatfield, we chose to
start the reduction analysis from the unflatfielded ({\it droopres})
images.  From there, the reduction and extraction techniques were
carried out as follows: Rogue pixels and flagged data were removed
using the {\em irsclean}\footnote{This tool is available from the SSC
  website: http://ssc.spitzer.caltech.edu} tool, which uses a mask of
rogue pixels for each campaign to first flag and then remove rogue
pixels.  These rogue pixels are especially troublesome in LH.  To
remove the contribution from the background in the low-resolution
modules (SL and LL) the images were differenced nod$-$nod (e.g.  SL1
nod 1$-$SL1 nod 2).  Spectra were extracted from the images using a
script version of the Cornell-developed software package SMART
\citep{hig}, using variable-column extraction for SL and LL, and
full-slit extraction for SH and LH.
The spectrum was calibrated by dividing the extracted spectrum of the
source by the extracted spectrum of a standard star (HR\,6348 for SL
and LL, and $\xi$\,Dra for SH and LH) and multiplying by its template
\citep[][ Sloan in prep]{coh}. Spikes which were not present in both
nod positions or in the overlap region between orders were treated as
artifacts and were removed manually.

It is important to note that there is a significant mismatch between
the different modules, probably due to the presence of another source
about 4\arcsec~from the target, which is revealed in the IRAC images
from the SAGE program \citep{mei}, and which is included in the LL and
LH slits.  Pointing effects may also be present. We used the
coordinates reported by \cite{lei}, which differ from the coordinates
recently reported from {\em HST} observations \citep{sha2} by
3\arcsec~in RA. We scaled the modules to match the average spectrum of
the two nod positions in SH, because the difference between the two
nods in SH is of only 5\% and one of the nod positions  avoids
the contaminating source.

\section{Analysis}

\begin{figure}
      \includegraphics[angle=0,width=7.5cm]{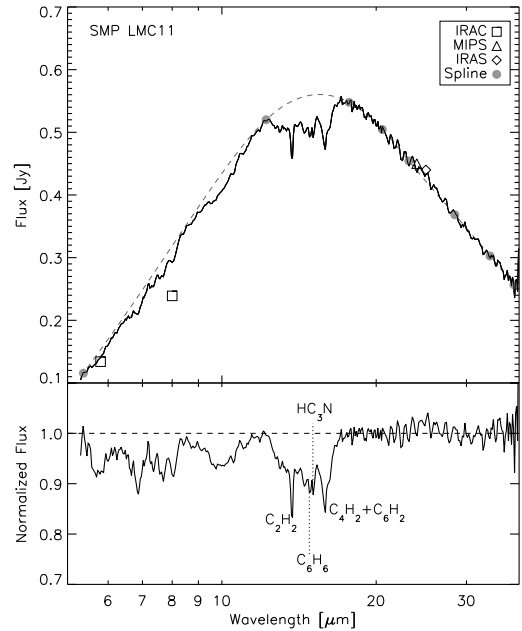}
    \caption{The low-resolution spectrum of SMP~LMC~11 (top). The
      dashed line indicates a spline fit representing the assumed
      continuum.  The photometric points correspond well with IRS
      spectrum except for the 8um IRAC which is slightly lower.  The
      bottom pannel shows the continuum-divided spectrum, revealing a
      wealth of absorption features due to various molecules (see
      Figure\,\ref{hr_f} and text for more details).\label{lr_f}}
  \end{figure}

Figure \ref{lr_f} (top) shows the resulting low-resolution spectrum of
SMP~LMC~11 together with the IRAC (5.8 and 8$\mu$m), MIPS (24 $\mu$m)
and IRAS (25 $\mu$m) fluxes.  A cool dust component dominates the
overall shape of the continuum emission.  The spectrum also shows
broad absorption bands, most notably between 12 and 17\,$\mu$m.

  \subsection{The dust continuum}

  The dashed line in Fig.\,\ref{lr_f} (top) is a spline fit to the
  dust continuum; the grey circles indicate the anchor points.  In the
  short wavelength region we anchored to the maximum emission at only
  two points, 5 and 12\,$\mu$m.  While this is an arbitrary choice,
  the resulting absorption spectrum is in accordance with the lack of
  emission features seen in the rest of the spectrum. Given the
  carbon-rich nature of the gas and the lack of broad emission
  features in the spectrum (such as PAHs), it is reasonable to assume
  that amorphous carbon grains dominate the dust mixture.  The dust
  continuum in Figure\,\ref{lr_f} is featureless and peaks at 15.5
  $\mu$m, implying that the bulk of the emitting dust is at 330\,K.
  For the remainder of this paper, we will represent the dust
  continuum by the spline fit shown in Figure\,\ref{lr_f}.

  \subsection{Absorption features}

  The bottom panel of Figure\,\ref{lr_f} shows the spectrum of SMP
  LMC\,11 after dividing out the dust continuum and reveals a number
  of molecular bands visible even at low resolution.  Similarly,
  Figure 2 shows the high-resolution spectrum of SMP LMC\,11 and the
  {\it ISO}/SWS spectrum of AFGL~618 for comparison.  In Figure 2, the
  narrow spectral features in SMP~LMC~11 are very similar to those in
  AFGL~618.

\begin{figure*}

  \begin{center}
  \includegraphics[height=8cm]{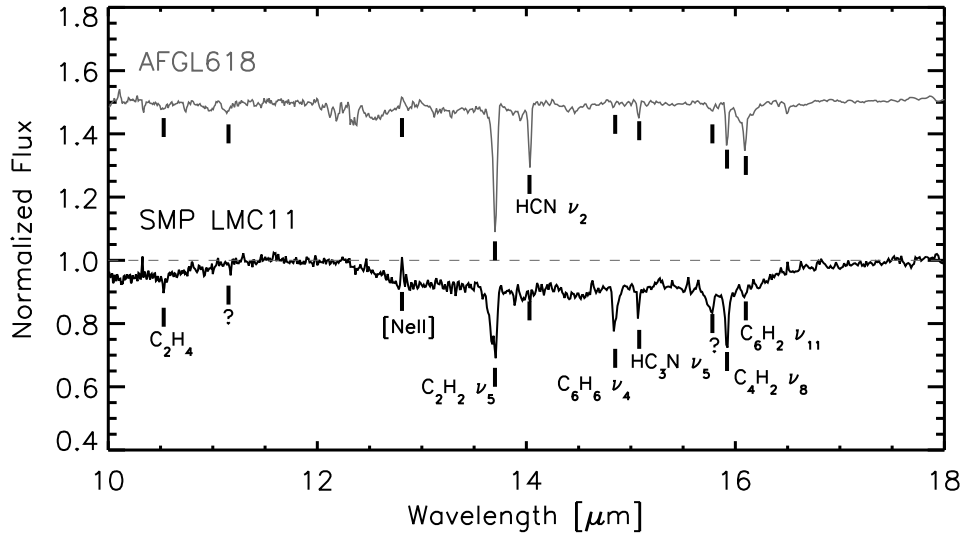}
  \caption{Normalized IRS high-resolution spectra (SH) of SMP~LMC~11,
    adopting the continuum from Figure\,\ref{lr_f} (bottom spectrum).
    For comparison the normalized ISO-SWS spectrum of AFGL~618 is also
    shown in grey with an offset of 0.5.  The spectrum of AFGL~618
    has been rebinned to the resolution of the IRS-SH module
    (R$\sim$600). Many molecular bands are clearly present in both
    spectra and some are only present in the spectrum of SMP LMC\,11.
    Q-branch bands are indicated with vertical tick marks with their
    identification if known.\label{hr_f}}
  \end{center}
  \end{figure*}

  The high-resolution spectrum of SMP~LMC~11 shows broad absorption
  between 12 and 17 \mic\ with a wealth of narrow lines superposed.
  These are the unresolved $Q$ branches of various ro-vibrational
  bands from different molecules, many of which can be identified by
  comparison to AFGL~618 as due to acetylene (C$_2$H$_2$,
  13.70\,$\mu$m), di-acetylene (C$_4$H$_2$, 15.87$\mu$m),
  tri-acetylene (C$_6$H$_2$, 16.10$\mu$m), benzene (C$_6$H$_6$,
  14.85$\mu$m) and HC$_3$N (15.10$\mu$m). Most of these also appear in
  the low-resolution spectrum (Figure \ref{lr_f}).  The strongest $Q$
  branch is the one due to C$_2$H$_2$ at 13.7\,$\mu$m.  Close
  inspection of the spectrum indicates the presence not only of the
  fundamental $\nu_5$ mode, but also of hot bands and combination
  bands.  The broad absorption is the result of the superposition of
  the many $P$- and $R$-branch lines that correspond to these $Q$
  branches, which indicates that the column densities of these species
  must be high. SMP~LMC~11 presents a weaker tri-acetylene band
  compared to AFGL~618, but in contrast shows a stronger benzene band.
  Ammonia (NH$_3$, with the strongest bands at 10.35, 10.75, and
  6.15\,$\mu$m) and HCN (at 14.01$\mu$m) are at most marginally
  present. By contrast stronger ro-vibrational bands of HCN are
  present in AFGL~618 and in the extreme carbon star IRC\,+10216
  \citep{cer99}. Figure\,\ref{lr_f} shows that the spectrum also
  exhibits an absorption dip near 10$\mu$m. Such a dip is also seen by
  \citet{zij06} in their sample of carbon stars.  They attribute this
  feature to C$_3$; however, this identification is still a matter of
  debate (e.g.  Speck et al. 2006).
 
  The spectrum of SMP~LMC~11 shows features which are not seen in
  AFGL~618 at 10.53, 11.15, and 15.78\,$\mu$m.  The feature at
  10.53\,$\mu$m corresponds to the $Q$-branch of the $\nu_7$
  CH$_2$-wagging mode of C$_2$H$_4$ (ethylene). In the {\it Spitzer}
  wavelength range, the only other mode of this molecule that would be
  expected is the strong scissoring mode at 6.93\,$\mu$m; the rocking
  mode at 12.10\,$\mu$m is expected to be weak. There is indeed a
  feature around 6.93 $\mu$m in the low-resolution spectrum
  (Figure\,\ref{lr_f}). CH$_3$C$_2$H, also known as propyne or
  methylacetylene, has its strongest band at 15.79\,$\mu$m with
  medium-strength bands at 6.84 and 7.97\,$\mu$m, and the spectrum of
  SMP~LMC~11 shows bands at both of those wavelengths (see Figure\,1).
  Therefore propyne may be a good candidate for the 15.78\,$\mu$m
  band.  However, the absorption from 6.3 to 8.3\,$\mu$m in Figure\,1
  (bottom panel) can also be due to the C$_2$H$_2$ $\nu_4$+$\nu_5$
  bands, which have been identified by \citet{mat} in a sample of LMC
  AGB stars.  With the latter identification the 8.0$\mu$m feature
  could be ascribed to the C$_4$H$_2$ $\nu_6$+$\nu_8$ band. The blue
  shoulder of this band in our spectrum may be due to HCN.
  \cite{slo06} noted that the two absorption bands from C$_2$H$_2$ at
  7.5 and 13.7\,$\mu$m are decoupled in their sample of carbons stars
  in the SMC.  Thus we cannot use the lack of a 14.0\,$\mu$m band from
  HCN to rule out a contribution of HCN at 7.1\,$\mu$m.  CH$_3$ at
  7.2\,$\mu$m and CH$_4$ at 7.7\,$\mu$m have been seen in AFGL~618
  \citep{cer01b} and in Titan's atmosphere \citep{cou}, and they could
  also be contributing to this band in SMP~LMC~11.  A firm
  identification of these bands will require higher resolution
  observations, especially from the ground and {\it JWST},
  complemented with a detailed comparison with laboratory
  measurements.  Besides the absorption features mentioned above, the
  spectrum of SMP~LMC~11 shows no traces of the MgS feature ($\sim$30
  $\mu$m) which is commonly seen in carbon stars with optically thick
  dust shells and on some PNe.

  \subsection{Emission lines}
  
  In addition to the absorption features, the high-resolution spectrum
  of SMP~LMC~11 clearly shows emission from \ion{[Ne}{2]} at 12.8
  \mic. The \ion{[Ne}{3]} line (15.5 \mic) may be present but
  the presence of another line at 15.65$\mu$m which may be a glitch
  makes the identification of the \ion{[Ne}{3]} more doubtful.  The
  presence of the low-excitation line of \ion{[Ne}{2]} is consistent
  with the optical measurements by \citet{mor} and \citet{sha2} of the
  low-excitation lines of \ion{[N}{2]}, \ion{[S}{2]}, \ion{[O}{1]} and
  \ion{[O}{3]}.

\section{Discussion}

The infrared spectrum of SMP~LMC~11 reveals three components: a cool
dust continuum, absorption bands from hydrocarbon molecules, and
emission from at least one low-ionization forbidden line.

The mean temperature of the dust as inferred from the peak of the IR
emission ($\sim$330~K) falls closer to the most evolved AGB stars
(T$\gtrsim$300~K) than the youngest PNe (T$\sim$150~K) \citep{kwo}.
The peak of the dust emission will shift to longer wavelengths as dust
moves further away from the central star.

The {\it Spitzer} spectrum of SMP~LMC~11 represents the first
extragalactic detection of several molecules, including di-acetylene
(C$_4$H$_2$), tri-acetylene (C$_4$H$_2$), benzene (C$_6$H$_6$),
HC$_3$N, C$_2$H$_4$, and possibly propyne.  These molecules are the
building blocks from which more complex hydrocarbons are produced.
The absorption bands are due to the increase in thickness of the
circumstellar shell, which is more dramatic at the very end of the AGB
phase \citep{gar}.  For free molecules, this absorption will not be a
Gaussian or a Lorentzian, but will be a complex addition of many
individual ro-vibrational lines with different strengths (which depend
on the temperature), causing the strong absorption we see in the
12-17\,$\mu$m region (due to the P and R branches of the molecules
present in that region).  The stronger absorption in this region
compared to AFGL~618 indicates that SMP~LMC~11 has a larger molecular
column density.

The spectrum of SMP~LMC~11 does not contain any PAH emission features.
A mixture of PAHs and the PAH-precursor C$_2$H$_2$ are seen together
in the low-resolution IRS spectrum of the post-AGB object MSX SMC\,029
\citep{kra}.  The absence of PAH emission in SMP~LMC~11 can be
attributed to age or orientation effects.  \cite{cer01a} propose that
PAHs may not form until the end of the AGB phase, based on their
observations of AFGL 618.  Alternatively, the dust in SMP~LMC~11 could
be distributed in an edge-on torus while MSX SMC\,029 is seen more
pole-on or has a more spherically symmetric dust distribution.  The
PAHs emit in photo-dissociation regions, and in an edge-on torus, we
would not have a clear line of sight to the emitting region. The dust
envelope is still too thick to allow a direct line of sight into the
hotter central regions.

Despite the detection of HC$_3$N (and perhaps HC$_5$N), the spectrum
of SMP~LMC~11 shows evidence for reduced abundances of nitrogen-based
molecules such as HCN and NH$_3$ compared to Galactic objects like
AFGL~618.  This agrees with the nitrogen abundance derived by
\cite{lei2} for SMP~LMC~11 ($\log(N/H)+12=7.1$) which is one of the
lowest in their sample of LMC PNe. This result fits with the finding
of \citet{mat} that the HCN bands that are prominent in the Galactic
sample of carbon stars are significantly weaker or absent in the LMC
sample.  The lower nitrogen abundance in the LMC compared to the
Galaxy and higher efficiency of the third dredge-up at lower
metallicities work together to decrease the nitrogen abundance
relative to carbon.

The presence of an optically thick and relatively warm dust component
between us and the central ionized region can explain not only the
absence of PAH features, but also the absence of most ionized lines in
the infrared.  We only see [Ne II] (and maybe \ion{[Ne}{3]}), and this
line is probably visible only in scattered light.  Scattering is more
efficient in the optical than in the infrared, which would explain why
more of the forbidden lines can be detected in the optical.  The dust
also explains the presence of molecular absorption in the spectrum,
since the molecules do not have a direct line of sight to the still
shielded central ionized region.  It is also possible that some of the
forbidden line emission is excited by high-velocity (v$>$100 km
s$^{-1}$) shocks \citep{lei2}.  All of these properties are more
consistent an object which is in transition from the AGB to a PN than
with a classic PN.

The bright H$\beta$ luminosity and its large size, present some
problems with this scenario of a transition object.  While the
measured H$\beta$ flux by \cite{sha2} makes SMP~LMC~11 one of the
faintest PNe in their LMC sample, its equivalent flux at 1 kpc (if we
assume that it is at a distance of 50 kpc) is log(H$\beta$)=$-$10.55,
which is bright for Galactic standards.  Similarly, \cite{sha2} state
that it is one of the smallest bipolar nebulae in their LMC sample,
but at a Galactic distance of 1 kpc it would be resolved. These
characteristics resemble those of a typical PN. One could invoke then
a massive molecular envelope surrounding the ionized region to explain
the absorption features.  Still, the absence of PAHs and H$_2$ in the
spectrum, and the high temperature of the dust (compared to PNe) would
remain unexplained (unless perhaps invoking a complex geometry).

The rich chemistry seen in the IRS spectrum of SMP~LMC~11 makes it
(together with AFGL~618, and MSX SMC\,029) a key object in our
understanding of the formation of hydrocarbons, deserving future study
from the community. Particularly, CO measurements of the circumstellar
AGB envelope may prove to be particularly useful.

\acknowledgments We would like to thank Richard Shaw for sharing his
results prior to publication, and the referee, whose diligence led to
substantial improvements in the manuscript. This work is based on
observations made with the Spitzer Space Telescope, which is operated
by the Jet Propulsion Laboratory, California Institute of Technology
under NASA contract 1407. Support for this work was provided by NASA
through Contract Number 1257184 issued by JPL/Caltech.



\begin{thebibliography}{}

\bibitem[Allamandola et al.(1989)]{atb89} Allamandola, L.~J., Tielens,
  A.~G.~G.~M., \& Barker, J.~R. 1989, \apjs, 71, 733

\bibitem[Asplund et al.(2005)]{asp} Asplund, M., Grevesse, N., \&
  Sauval, A.J. 2005, ASPC, 336, 25

\bibitem[Beichman et al.(1988)]{bei} Beichman, C.A., Neugebauer, G.,
  Harbig, H.J., Clegg, P.E., \& Chester, T.J. 1988, NASA RP-1990,
  Infrared Astronomical Satellite (IRAS) Catalogues and Atlases, Vol
  1: Explanatory Supplement

\bibitem[Blommaert et al.(2005)]{Blommaert:ISOlegacy} Blommaert J.,
  Cami J., Szczerba R. and Barlow M.J. 2005, in ``ISO science legacy -
  a compact review of ISO major achievements'', Space Science Reviews,
  119, 215

\bibitem[Bujarrabal et al.(1988)]{buj} Bujarrabal, V.,
  G\'omez-Gonz\'alez, J., Bachiller, R., \& Mart\'in-Pintado, J.,
  1988, A\&A, 204, 242

\bibitem[Cernicharo et al.(1999)]{cer99} Cernicharo, J., Yamamura, I.,
  Gonz\'alez-Alonso, E., et al. 1999, \apj, 526, L41

\bibitem[Cernicharo et al.(2001a)]{cer01a} Cernicharo, J., Heras, A.M.,
  Tielens, A.G.G.M., et al. 2001a, \apj, 546, L123

\bibitem[Cenicharo et al.(2001b)]{cer01b} Cernicharo, J., Heras, A.M.,
  Pardo, J.R., et al. 2001b, \apj, 546, L127

\bibitem[Chiar et al.(1998)]{chi98} Chiar, J.E., Pendleton, Y.J.,
  Geballe, T.R., \& Tielens, A.G.G.M. 1998, \apj, 507, 281


\bibitem[Cohen et al.(2003)]{coh} Cohen, M., Megeath, T.G.,
  Hammersley, P. L., Martin-Luis, F., \& Stauffer, J. 2003, \aj, 125,
  2645
 
\bibitem[Coustenis et al.(2003)]{cou} Coustenis, A., Salama, A.,
  Schulz, B., et al. 2003, Icarus, 161, 383

\bibitem[Frenklach \& Feigelson(1989)]{ff89} Frenklach, M. \&
  Feigelson, E.~D. 1989, \apj, 341, 372

\bibitem[Garc\'{i}a-Lario \& Perea Calder\'on(2002)]{gar}
  Garc\'{i}a-Lario, G., Perea Calder\'on, J.V. 2002, ESA SP-511,
  Proceedings of the Symp. Exploiting the ISO Data Archive - Infrared
  Astronomy in the Internet Age, pg 97

\bibitem[de Graauw et al.(1996)]{deg}
de Graauw, T., Haser, L.N., Beintema, D.A., et al., 1996, A\&A 315, L49

\bibitem[Habing (1996)]{Habing96}Habing, H.J. 1996, A\&AR, 7, 97

\bibitem[Higdon et al.(2004)]{hig} Higdon, S.J.U., Devost, D., Higdon,
  J.L. et al. 2004, PASP, 116, 975

\bibitem[Houck et al.(2004)]{hou} Houck, J. R., Appleton, P. N.,
  Armus, L., et al. 2004, \apjs, 154, 18

\bibitem[Kessler et al.(1996)]{kes} Kessler, M.F., Steinz, J.A.,
  Anderegg, M.E., et al. 1996, A\&A, 315, L27

\bibitem[Kraemer et al.(2006)]{kra} Kraemer, K.E., Sloan, G.C.,
  Bernard-Salas, J., et al. 2006, ApJ, submitted


\bibitem[Kwok(2000)]{kwo} Kwok, S. 2000, Cambridge Astrophysics Series 31,
  The Origin and Evolution of Planetary Nebulae

\bibitem[Leisy \& Dennefeld(2006)]{lei2} Leisy, P., \& Dennefeld, M.
  2006, A\&A, in press

\bibitem[Leisy et al.(1997)]{lei} Leisy, P., Dennefeld, M., Alard, C.,
  \& Guibert, J. 1997, \aaps, 121, 407

\bibitem[Matsuura et al.(2006)]{mat} Matsuura, M., Wood, P., Sloan,
  G.C., et al. 2006, MNRAS, 371, 415

\bibitem[Meatheringham and Dopita(1991a)]{mea91a} Meatheringham, S.
  J., \& Dopita, M. A. 1991, \apjs, 75, 407

\bibitem[Meatheringham and Dopita(1991b)]{mea91b} Meatheringham, S.
  J., \& Dopita, M. A. 1991, \apjs, 76, 1085

\bibitem[Meatheringham et al.(1988)]{mea} Meatheringham, S.J., Dopita,
  M.A., \& Morgan, D.H. 1988, \apj, 329, 166

\bibitem[Meixner et al.(2006)]{mei} Meixner, M., Gordon, K.D.,
  Indebetouw, R., et al. 2006, \aj, in press

\bibitem[Morgan \& Parker(1998)]{mor} Morgan, \& Parker 1998, \mnras,
  296, 921

\bibitem[Sanduleak et al.(1978)]{san} Sanduleak, N., MacConnel, D.J.,
  \& Davis Philip, A.G. 1978, PASP, 90, 621

\bibitem[Sloan et al.(2006)]{slo06} Sloan, G.C., Kraemer, K.E.,
  Matsuura, M., et al. 2006, \apj, 645, 1118

\bibitem[Shaw et al.(2006)]{sha2} Shaw, R.A., Stanghellini, L.,
  Villaver, E., \& Mutcher, M. 2006, \apj, submitted


\bibitem[Speck et al.(2006)]{spe} Speck, A.K., Cami, J.,
  Markwick-Kemper, C., et al. 2006, ApJ, in press

\bibitem[Werner et al.(2004)]{wer} Werner, M., Roellig, T. L., Low, F.
  J., et al. 2004, \apjs, 154, 1

\bibitem[Wood et al.(1987)]{woo} Wood, P.R., Meatheringham, S.J.,
  Dopita, M.A., and Morgan, D.H. 1987, \apj, 320, 178

\bibitem[Zijlstra et al.(2006)]{zij06} Zijlstra, A.A., Matsuura, M.,
  Wood, P., et al. 2006, MNRAS, 370, 1961


\end{thebibliography}
\end{document}